\documentclass[aps,prb,floatfix,amsmath,amssymb,preprint,eqsecnum]{revtex4}
\usepackage{graphicx}
\usepackage{dcolumn}
\usepackage{bm}
\usepackage{epstopdf}
\usepackage{enumerate}
\usepackage{setspace}   
\usepackage{amsmath}
\usepackage{amssymb}

\preprint{arXiv:0905.2608}

\begin{document}
\title{Competition between spin density wave order and superconductivity in the underdoped cuprates}

\author{Eun Gook Moon}
\affiliation{Department of Physics, Harvard University, Cambridge MA
02138}

\author{Subir Sachdev}
\affiliation{Department of Physics, Harvard University, Cambridge MA
02138}

\date{May 14, 2009\\
\vspace{1.6in}}
\begin{abstract}
We describe the interplay between $d$-wave superconductivity
and spin density wave (SDW) order in a theory of the hole-doped cuprates at hole densities below
optimal doping.
The theory assumes local SDW order, and associated electron
and hole pocket Fermi surfaces of charge carriers in the normal state. We describe  
quantum and thermal fluctuations in the orientation of the local SDW order, which lead
to $d$-wave superconductivity: we compute the superconducting 
critical temperature and magnetic field in a `minimal' universal theory.
We also describe the back-action
of the superconductivity on the SDW order, showing that SDW order is more stable in the metal. 
Our results capture key aspects of 
the phase diagram of Demler {\em et al.} (Phys. Rev. Lett. {\bf 87}, 067202 (2001))
obtained in a phenomenological quantum theory
of competing orders. Finally, we propose a finite temperature crossover phase diagram for 
the cuprates. In the metallic state, these are controlled by a `hidden' quantum critical point
near optimal doping involving the onset of SDW order in a metal. However, the onset of 
superconductivity results in a decrease in stability of the SDW order, and consequently the actual
SDW quantum critical point appears at a significantly lower doping.
 All our analysis is placed in the context
of recent experimental results. 
\end{abstract}

\maketitle

\section{Introduction}
\label{sec:intro}

A number of recent experimental observations have the potential to dramatically
advance our understanding of the enigmatic underdoped regime of the cuprates.
In the present paper, we will focus in particular on two classes of experiments (although
our results will also have implications for a number of other experiments):
\begin{itemize}
\item
The observation of quantum oscillations in the underdoped region of YBCO. \cite{doiron,cooper,nigel,cyril,suchitra,louis}
The period of the oscillations implies a carrier density of order the density of dopants.
LeBoeuf {\em et al.} \cite{louis} have claimed that the oscillations are actually due
to electron-like carriers of charge $-e$. We will accept this claim here,
and show following earlier work \cite{rkk3,gs}, that it helps resolve a number of
other theoretical puzzles in the underdoped regime.
\item Application of a magnetic field to the superconductor induces a quantum phase transition at
a non-zero critical field, $H_{\rm sdw}$, involving the onset of spin density wave (SDW) order. This transition was
first observed in La$_{2-x}$ Sr$_x$CuO$_4$ with $x=0.144$ 
by Khaykovich {\em et al.} \cite{boris}. Chang {\em et al.} \cite{mesot,mesot2}
have provided detailed studies of the spin dynamics in the vicinity of $H_{\rm sdw}$, including observation
of a gapped spin collective mode for $H<H_{\rm sdw}$ whose gap vanishes as $H \nearrow H_{\rm sdw}$.
Most recently, such observations have been extended to YBa$_2$Cu$_3$O$_{6.45}$ by Haug {\em et al.} \cite{mesot3}, who obtained
evidence for the onset of SDW order at $H \approx 15$ T. These observations were all on systems which do
not have SDW order at $H=0$; they build on the earlier work of Lake {\em et al.} \cite{lake} who observed enhancement
of prexisting SDW order at $H=0$ by an applied field in La$_{2-x}$ Sr$_x$CuO$_4$ with $x=0.10$.
\end{itemize}

\begin{figure}
\includegraphics[width=4.0in]{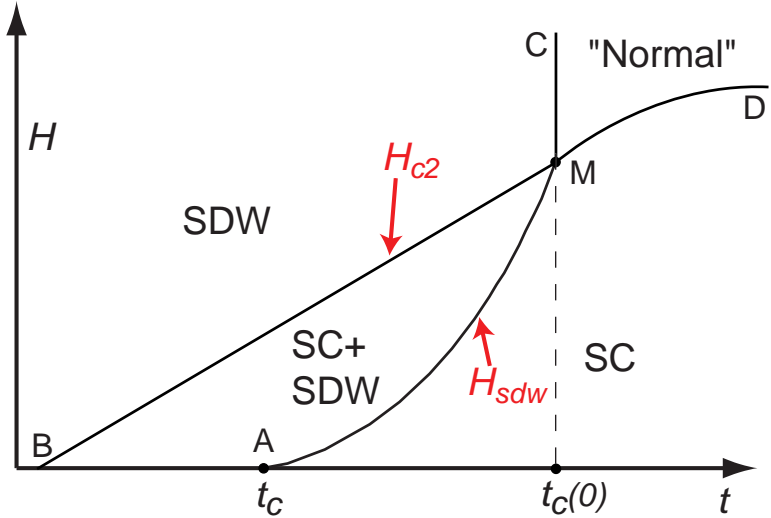}
\caption{From Ref.~\onlinecite{demler}: Phase diagram of the competition between
superconductivity (SC) and spin density wave (SDW) order tuned by an applied magnetic field $H$, and a Landau parameter $t$ controlling the SDW order (the effective action has a term $t \vec{\varphi}^2$, where
$\vec{\varphi}$ is the SDW order). The labels identifying $H_{c2}$, $H_{\rm sdw}$, and $t_c (0)$ have been
added to the original figure, \cite{demler} but the figure is otherwise unchanged. 
The dashed line does not indicate any transition or crossover; it is just
the continuation of the line CM to identify $t_c (0)$.
A key feature of this phase
diagram is that SDW order is more stable in the metal than in the superconductor {\em i.e.\/} $t_c (0) > t_c$.
}
\label{figdemler}
\end{figure}

We begin our discussion of these experiments using the phenomenological quantum theory of the competition
between superconductivity and SDW order.\cite{demler,ssrmp,kivelsonlee,stripes}
The phase diagram in the work of Demler {\em et al.}\cite{demler} is reproduced in Fig.~\ref{figdemler}.
The parameter $t$
appears in a Landau theory of SDW order and tunes the propensity to SDW order, with SDW order
being favored with decreasing $t$. We highlight a number of notable features of this phase diagram:
\begin{enumerate}[~~~A.]
\item The upper-critical field above which superconductivity is lost, $H_{c2}$, decreases with decreasing
$t$. This is consistent with the picture of competing orders, as decreasing $t$ enhances the SDW order,
which in turn weakens the superconductivity.
\item The SDW order is more stable in the non-superconducting `normal' state than in the superconductor.
In other words, the line CM, indicating the onset of SDW order in the normal state, is to the
right of the point A where SDW order appears in the superconductor at zero field; {\em i.e.\/} $t_c (0)
> t_c$. Thus inducing
superconductivity destabilizes the SDW order, again as expected in a model of competing orders.
\item An immediate consequence of the feature B is the existence of the line AM of quantum
phase transitions within the superconductor, representing $H_{\rm sdw}$, where SDW order appears
with increasing $H$. As we have discussed above, this prediction of Demler {\em et al.}\cite{demler}
has been verified in a number of experiments.
\end{enumerate}
A related prediction by Demler {\em et al.} \cite{demler} that an applied current should enhance the SDW
order, also appears to have been observed in a recent muon spin relaxation experiment.\cite{amit}

\begin{figure}
\includegraphics[width=6.3in]{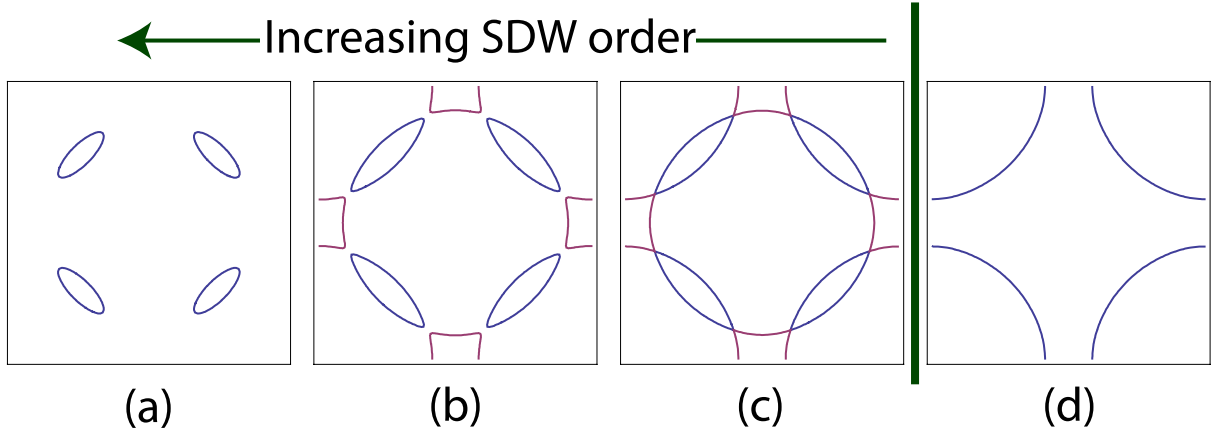}
\caption{(Color online) Fermi surface evolution in the SDW theory \cite{sokol,morr}. Panel (d) is the ``large Fermi surface''
state appropriate for the overdoped superconductor. The SDW order parameter, $\vec{\varphi}$, desribes ordering at the wavevector ${\bf Q} = (\pi,\pi)$, and mixes fermion states whose wavevectors differ
by ${\bf Q}$. This leads to the SDW metal state with electron (red) and hole (blue) pockets in panel (b),
which is the state used here to explain the quantum oscillation experiments.\cite{doiron,cooper,nigel,cyril,suchitra,louis} }
\label{figscs}
\end{figure}

A glance at Fig.~\ref{figdemler} shows that it is natural to place \cite{fuchun} the quantum oscillation experiments \cite{doiron,cooper,nigel,cyril,suchitra,louis} in the non-superconducting phase labeled ``SDW''. Feature B
above is crucial in this identification: the normal state reached by suppressing superconductivity with a field is a regime
where SDW order is more stable. The structure of the Fermi surface in this normal state can be deduced
in the framework of conventional spin-density-wave theory, and we recall the early results of Refs.~\onlinecite{sokol,morr} in Fig.~\ref{figscs}.
Recent studies \cite{millisnorman,harrison} have extended these results to incommensurate
ordering wavevectors ${\bf Q}$, and find that the electron pockets (needed to explain the quantum oscillation experiments) remain robust under deviations from the commensurate ordering at $(\pi, \pi)$.
The present paper will consider only the case of commensurate ordering with ${\bf Q} = (\pi,\pi)$,
as this avoids considerable additional complexity.

The above phenomenological theory appears to provide a satisfactory framework for interpreting the
experiments highlighted in this paper. However, such a theory cannot ultimately be correct.
A sign of this is that within its parameter space is a non-superconducting,
non-SDW normal state at $H=0$ and $T=0$ (not shown in Fig.~\ref{figdemler}). 
Indeed, such a state is the point of departure for
describing the onset of the superconducting and SDW order in  Ref.~\onlinecite{demler}. 
There is no such physically plausible state,
and the parameters were chosen so that this state does not appear in Fig.~\ref{figdemler}.
Furthermore, we would like to extend the theory to
spectral properties of the electronic excitations probed in numerous other experiments.
This 
requires a more microscopic formulation of the theory of competing orders in terms of the underlying electrons. We shall provide such a theory here, building upon the proposals of Refs.~\onlinecite{rkk3,gs,rkk1,rkk2}. 
Our theory  
will not have the problematic $H=0$, $T=0$ ``normal'' state of the phenomenological theory, and so cannot be mapped precisely onto it. Nevertheless, we will see
that our theory does reproduce the key aspects of Fig.~\ref{figdemler}.
We will also use our theory to propose a finite temperature phase diagram for the hole-doped cuprates; in particular, we will argue
that it helps resolve a central puzzle on the location of the quantum critical point important for the finite temperature crossovers
into the `strange metal' phase. These results appear in Section~\ref{sec:conc} and Fig.~\ref{figT}.

The theory of superconductivity \cite{dwave} mediated by exchange of quanta
of the SDW order parameter, $\vec{\varphi}$, has been successful above optimal doping.
However, it
does not appear to be compatible with the physics
of competing orders in the underdoped regime, at least in its simplest version. 
This theory begins with the ``large Fermi surface'' state
in panel (d) of Fig.~\ref{figscs}, and examines its instability in a BCS/Eliashberg theory due to attraction
mediated by exchange of $\vec{\varphi}$ quanta. An increase in the fluctuations of $\vec{\varphi}$ is
therefore connected to an increase in the effective attraction, and consequently a strengthening of the
superconducting order. This is evident from the increase in the critical temperature for superconductivity
as the SDW ordering transition is approached from the overdoped side (see {\em e.g.\/} Fig.~4 in Ref.~\onlinecite{abanov}). Thus rather than a competition, this theory yields an effective attraction
between the SDW and superconducting order parameters. This was also demonstrated in Ref.~\onlinecite{demler} by a microscopic computation in this framework of the coupling between these order
parameters. It is possible that these difficulties may be circumvented in more complex strong-coupling versions
of this theory \cite{abanov}, but a simple physical picture of these is lacking.  

As was already discussed in Ref.~\onlinecite{demler}, the missing ingredient in the SDW theory of the ordering
of the metal is the knowledge of the proximity to the Mott insulator in the underdoped compounds. Numerical studies of models in which the strong local repulsion associated with Mott insulator is implemented in a mean-field manner do appear to restore aspects of the picture of competing orders. \cite{kwon,trivedi}
Here, we shall provide a detailed study of the model of the underdoped cuprates proposed
in Refs.~\onlinecite{rkk3,gs,rkk1,rkk2}, and show that it is consistent with the features A, B, and C of the theory of
competing orders noted above, which are essential in the interpretation of the experiments.

As discussed at some length in Ref.~\onlinecite{gs}, the driving force of the superconductivity in the underdoped
regime is argued to be the pairing of the electron pockets visible in panel (b) of Fig.~\ref{figscs}. Experimental evidence for
this proposal also appeared in the recent photoemission experiments of Yang {\em et al.} \cite{yang}. In the interests of simplicity,
this paper will focus exclusively on the electron pockets, and neglect the effects of the hole pockets in Fig.~\ref{figscs}.
Further discussion on the hole pockets, and the reason for their secondary role in superconductivity
may be found in Refs.~\onlinecite{gs,rkk1,rkk2}.

The degrees of freedom of the theory are the bosonic spinons $z_\alpha$ ($\alpha = \uparrow, \downarrow$),
and spinless fermions $g_\pm$. The spinons determine the local orientation of the SDW order via
\begin{equation}
\vec{\varphi} = z_\alpha^\ast \vec{\sigma}_{\alpha\beta} z_\beta \label{pz}
\end{equation}
where $\vec{\sigma}$ are the Pauli matrices. The electrons are assumed to form electron and hole pockets as indicated
in Fig.~\ref{figscs}b, but with their components determined in a `rotating reference frame' set 
by the local orientation of $\vec{\varphi}$. This idea of Fermi surfaces correlated with the local order is supported by
the recent STM observations of Wise {\em et al.}\cite{eric2}.
Focussing only on the electron pocket components, we can write the physical electron operators $c_\alpha$ as \cite{rkk3,gs}
\begin{eqnarray}
c_\uparrow &=& e^{i {\bf G}_1 \cdot {\bf r}} \Bigl[ z_\uparrow g_+ - z_\downarrow^\ast g_- \Bigr]
+ e^{i {\bf G}_2 \cdot {\bf r}} \Bigl[  z_\uparrow g_+ + z_\downarrow^\ast g_-
\Bigr] \nonumber \\
c_\downarrow &=& e^{i {\bf G}_1 \cdot {\bf r}} \Bigl[   z_\downarrow g_+ + z_\uparrow^\ast g_-
\Bigr] + e^{i {\bf G}_2 \cdot {\bf r}} \Bigl[ z_\downarrow g_+ - z_\uparrow^\ast g_-
\Bigr] \label{Pgz}
\end{eqnarray}
where ${\bf G}_1 = (0, \pi)$ and ${\bf G}_2 = (\pi, 0)$ are the anti-nodal points about which the electron pockets are centered.
We present an alternative derivation of this fundamental relation from spin-density-wave theory in Appendix~\ref{sdw}.

Note that when $z_\alpha = (1,0)$, Eq.~(\ref{pz}) shows that
the SDW order is uniformly polarized in the $z$ direction with $\vec{\varphi} = (0,0,1)$,
and from Eq.~(\ref{Pgz}) we have $c_\uparrow = g_+ (e^{i {\bf G}_1 \cdot {\bf r}} + e^{i {\bf G}_2 \cdot {\bf r}} ) $ and
$c_\downarrow = g_- (e^{i {\bf G}_1 \cdot {\bf r}} - e^{i {\bf G}_2 \cdot {\bf r}})$. Thus, for this SDW state, the $\pm$ labels
on the $g_\pm$ are equivalent to the $z$ spin projection, and the spatial dependence is the consequence of the potential
created by the SDW order, which has opposite signs for the two spin components (as shown in Appendix~\ref{sdw}). The expression in Eq.~(\ref{Pgz}) for
general $\vec{\varphi}$ is then obtained by performing a spacetime-dependent spin rotation, determined by $z_\alpha$, on this reference state.

Another crucial feature of Eqs.~(\ref{pz}) and (\ref{Pgz}) is that the physical observables $\vec{\varphi}$ and $c_\alpha$
are invariant under the following U(1) gauge transformation of the dynamical variables $z_\alpha$ and $g_\pm$:
\begin{equation}
z_\alpha \rightarrow e^{i \phi} z_\alpha~~~;~~~g_+ \rightarrow e^{-i \phi} g_+~~~;~~~g_- \rightarrow e^{i \phi} g_-  .
\end{equation}
Thus the $\pm$ label on the $g_\pm$ can also be interpreted as the charge under this gauge transformation. This gauge invariance
implies that the low energy effective theory will also include an emergent U(1) gauge field $A_\mu$.

We will carry out most of the computations in this paper using a ``minimal model'' for $z_\alpha$ and $g_\pm$ with
the imaginary time ($\tau$) Lagrangian \cite{rkk3,gs}
\begin{equation}
\mathcal{L} = \mathcal{L}_z + \mathcal{L}_g , \label{m1}
\end{equation}
where the fermion action is
\begin{eqnarray}
\mathcal{L}_g &=& g_+^\dagger \left[ (\partial_\tau - i A_\tau )  - \frac{1}{2m^*} ( {\bm \nabla} - i {\bf A}  )^2 - \mu  \right] g_+
\nonumber \\ &+& g_-^\dagger \left[ (\partial_\tau + i A_\tau )  - \frac{1}{2m^*} ({\bm  \nabla} + i {\bf A} )^2 - \mu \right] g_-
, \label{lg}
\end{eqnarray}
and the spinon action is
\begin{equation}
\mathcal{L}_z = \frac{1}{t} \left[ \sum_{\alpha=1}^{N} \biggl(|(\partial_\tau - i A_\tau) z_\alpha|^2 +
v^2 |({\bm \nabla} - i {\bf A}) z_\alpha|^2 \biggr) + i \varrho \left( \sum_{\alpha=1}^N |z_\alpha|^2 - N \right) \right]. \label{lz}
\end{equation}
Here the emergent gauge field is $A_\mu = (A_\tau, {\bf A})$, and, for future convenience, we have generalized to
a theory with $N$ spin components (the physical case is $N=2$). The field $\varrho$ imposes a fixed length constraint
on the $z_\alpha$, and accounts for the self-interactions between the spinons.

This effective theory omits numerous other couplings involving higher powers or gradients of the fields, which have
been discussed in some detail in previous work. \cite{rkk1,rkk2,rkk3,gs} It also omits the $1/r$ Coulomb repulsion between the $g_\pm$ fermions--this
will be screened by the Fermi surface excitations, and is expected to reduce the critical temperature as in the 
traditional strong-coupling theory of superconductivity. For simplicity, we will neglect such effects here, as they are
not expected to modify our main conclusions on the theory of competing orders. Non-perturbative effects of Berry phases
are expected to be important in the superconducting phase, and were discussed earlier; \cite{rkk3} they should not be important
for the instabilities towards superconductivity discussed here.

As has been discussed earlier,\cite{rkk3,gs} the theory in Eq.~(\ref{m1}) has a superconducting ground state with a 
a simple momentum-independent pairing of the $g_\pm$ fermions $\langle g_+ g_- \rangle \neq 0$.
Combining this pairing amplitude with Eq.~(\ref{Pgz}), it is then easy to see \cite{rkk3,gs} that the physical
$c_\alpha$ fermions have the needed $d$-wave pairing signature (see Appendix~\ref{sdw}).

The primary purpose of this paper is to demonstrate that the simple field theory in Eq.~(\ref{m1}) satisfies the
constraints imposed by the framework of the picture of competing orders. In particular, we will show that it displays
the features A, B, and C listed above. Thus, we believe, it offers an attractive and unified framework for understanding
a large variety of experiments in the underdoped cuprates. We also note that the competing order interpretation of 
Eq.~(\ref{m1}) only relies on the general gauge structure of theory, and not specifically on the interpretation of $g_\pm$
as electron pockets in the anti-nodal region; thus it could also apply in other physical contexts.

Initially, it might seem that the simplest route to understanding the phase diagram of our theory Eq.~(\ref{m1}) is to use
it to compute the effective coupling constants in the phenomenological theory of Ref.~\onlinecite{demler}. However, such a literal
mapping is not possible, because, as we discussed earlier, the phenomenological theory does have additional unphysical
phases. Rather, we will show that our theory does satisfy the key requirements of the experimentally relevant
phase diagram in Fig.~\ref{figdemler}.

A notable feature of the theory in Eq.~(\ref{m1}) is that it is characterized by only 2 dimensionless couplings. We assume the chemical
potential $\mu$ is adjusted to obtain the required fermion density, which we determine by the value of the Fermi wavevector
$k_F$. The effective fermion mass $m^\ast$ and the spin-wave velocity then determine our first dimensionless ratio
\begin{equation}
\alpha_1 \equiv \frac{\hbar k_F}{m^\ast v}.
\end{equation}
Although we have inserted an explicit factor of $\hbar$ above, we will set $\hbar=k_B=1$ in most of our analysis.
Note that we can also convert this ratio to that of the Fermi energy, $E_F = \hbar^2 k_F^2 /(2 m^\ast)$ and the energy scale $m^\ast v^2$:
\begin{equation}
 \frac{E_F}{m^\ast v^2} = \frac{\alpha_1^2}{2}
\end{equation}
From the values quoted in the quantum oscillation experiment \cite{doiron}, $m^\ast = 1.9 m_e$ and $\pi k_F^2 = 5.1$~nm$^{-2}$, and the spin-wave velocity in the insulator $v \approx 670$ meV \AA, we obtain the estimate $\alpha_1 \approx 0.76$. We will also use 
\begin{equation}
m^\ast v^2 \approx 112~{\rm meV} 
\end{equation}
as a reference energy scale.

The second dimensionless coupling controls the strength of the fluctuations of the SDW order, which are controlled by
the parameter $t$ in Eq.~(\ref{lz}). Tuning this coupling leads to a transition from a phase with $\langle z_\alpha \rangle \neq 0$
to one where the spin rotation symmetry is preserved. We assume that this transition occurs at the value $t=t_c (0)$ in the metallic
phase (the significance of the argument of $t_c$ will become clear below): this corresponds to the line CM in Fig.~\ref{figdemler}. Then we can charaterize the deviation from this quantum phase
transition by the coupling
\begin{equation}
\alpha_2 \equiv \left( \frac{1}{t_c (0)} - \frac{1}{t} \right) \frac{1}{m^\ast}. \label{alpha2}
\end{equation}
Note that $\alpha_2 < 0$ corresponds to the SDW phase in Fig.~\ref{figdemler}, while $\alpha_2 > 0$
corresponds to the ``Normal'' phase of Fig.~\ref{figdemler}. For $\alpha_2>0$, we can also characterize
this coupling by the value of the spinon energy gap $\Delta_z$ in the $N=\infty$ theory, which is (as will become clear below)
\begin{equation}
\frac{\Delta_z}{m^\ast v^2} = 4 \pi \alpha_2 .
\end{equation}

It is worth noting here that our ``minimal model'' (Eq.~(\ref{m1})) in two spatial dimensions has aspects of the universal physics of the 
Fermi gas at unitarity in three spatial dimensions. The latter model has a `detuning' parameter which
tunes the system away from the Feshbach resonance; this is the analog of our parameter $\alpha_2$. The overall
energy scale is set in the unitary Fermi gas by the Fermi energy; here, instead, we have 2 energy scales, $E_F$
and $m^\ast v^2$.

The outline of the remainder of the paper is as follows. In Section~\ref{sec:eliash}, we will consider the pairing
problem of the $g_\pm$ fermions, induced by exchange of the gauge boson $A_\mu$. We will do this within
a conventional Eliashberg framework. Our main result will be a computation of the critical field $H_{c2}$,
which will be shown to be suppressed as SDW order is enhanced with decreasing $t$. Section~\ref{sec:shift} will consider
the feedback of the superconductivity on the SDW ordering, where we will find enhanced stability of the SDW
order in the metal over the superconductor. Section~\ref{sec:conc} will summarize our results, and propose a
crossover phase diagram at non-zero temperatures.

\section{Eliashberg theory of pairing}
\label{sec:eliash}

In our mininal model, the charge and spin excitations interact with each other through the $A_\mu$ gauge boson. So the gauge fluctuation is one of the key ingredients in our analysis. We begin by computing the gauge propagator, and  
then we will determine the critical temperature and magnetic field within the Eliashberg theory in the following subsections.

\begin{figure}
\includegraphics[width=4.0 in]{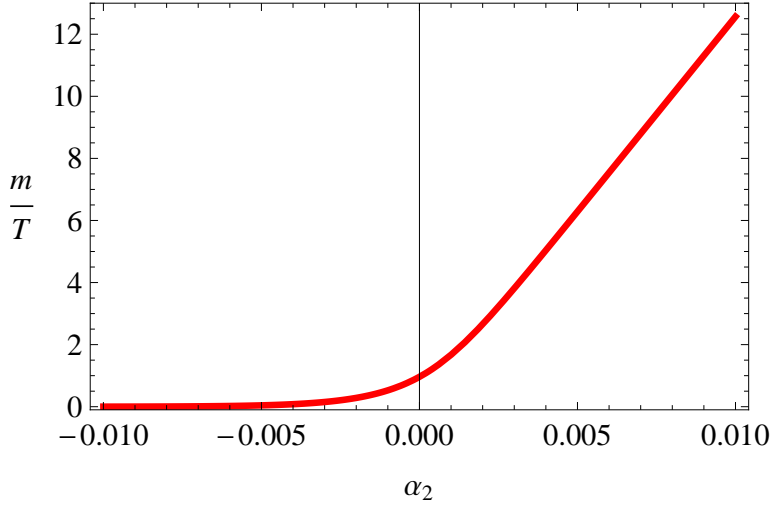}
\caption{The parameter $m$ in Eq.~(\ref{zprop}) for ${T}/{(m^\ast v^2)}=0.01$. }
\label{mass}
\end{figure}

We use the framework of the large $N$ expansion.  
In the limit $N = \infty $, the gauge field is suppressed, and the constraint field $\varrho$ takes a saddle point value ($i\varrho = m^2$) 
that makes the spinon action extremum in Eq.~(\ref{lz}).
At leading order, the spinon propagator has the form
\begin{equation}
\frac{t}{v^2 k^2 + \omega_n^2 + m^2}
\label{zprop}
\end{equation}
where $k$ is spatial momentum, $\omega_n$ is the Matsubara frequency.
The saddle point equation for $m$ is
\begin{equation}
T \sum_{\omega_n} \int \frac{d^2 k}{4 \pi^2} \left[  \frac{1}{v^2 k^2 + \omega_n^2 + m^2}  \right] = -m^{\ast} \alpha_{2} +  
\int \frac{d \omega}{2 \pi} \int  \frac{d^2 k}{4 \pi^2} \frac{1}{v^2 k^2 + \omega^2}.
\label{const}
\end{equation}
The solution of this is
\begin{equation}
m = 2 T \ln \left[ \frac{ e^{+2 \pi m^{\ast} v^2 \alpha_{2}/T} + \sqrt{ e^{+4 \pi m^{\ast}v^2 \alpha_{2}/T} + 4}}{2} \right]
\end{equation}
which holds for $-\infty < \alpha_{2} < \infty$.
This result is plotted in Fig.~\ref{mass}.
Clearly, $m$ is a monotonically increasing function of $\alpha_2$. Recall that the positive $\alpha_{2}$ region has no
SDW order, and $m$ is large here.
As we will see below, the value of $m$ plays a significant role in the photon propagators.

The photon propagator is determined from the effective action
obtained by integrating out the spinons and non-relativistic fermions.
Using gauge invariance, we can write down the effective action of the gauge field as follows:
\begin{equation}
S_A = \frac{NT}{2} \sum_{\epsilon_n} \int \frac{d^2 k}{4 \pi^2}
\left[ \left( k_i A_\tau - \epsilon_n A_i \right)^2 \frac{D_{1} (k,
\epsilon_n)}{k^2} + A_i A_j \left( \delta_{ij} - \frac{k_i k_j}{k^2}
\right) D_{2} (k, \epsilon_n) \right]. \label{sa}
\end{equation}
As in analogous computation with relativistic fermions in 
Ref.~\onlinecite{ribhu}, we separate the photon polarizations into
their bosonic and fermionic components:
\begin{eqnarray}
D_1 &=& N D_{1b} + D_{1f} \nonumber \\
D_2 &=& N D_{2b} + D_{2f}.
\end{eqnarray}
We use the Coulomb gauge, ${\bf k} \cdot {\bf A}=0$ in the
computation. After imposing the gauge condition, the propagator of
$A_\tau$ from the above action is $1/D_1$, while that of $A_i$ is
\begin{equation}
\left( \delta_{ij} - \frac{k_i k_j}{k^2} \right) \frac{1}{D_{2} +
(\epsilon_n^2/k^2) D_{1}}.
\end{equation}

We will approximate $D_{1b}$ and $D_{2b}$ by their zero frequency limits. Computation of the spinon polarization in this
limit, as in Ref.~\onlinecite{ribhu} yields
\begin{eqnarray}
D_{1b} (k) &=& - \frac{T}{\pi v^2} \ln \left( 2 \sinh\left(\frac{m}{2T} \right) \right)  \nonumber \\
&~&~~~ + \frac{1}{2 \pi v^2} \int_0^1 dx \sqrt{ m^2 + v^2 k^2 x (1-x)} \coth \left( \frac{\sqrt{ m^2 + v^2 k^2 x (1-x)}}{2 T} \right)
\end{eqnarray}
and
\begin{equation}
D_{2b} (k) =  \frac{v^2 k^2}{8 \pi} \int_0^1 dx \frac{1}{\sqrt{ m^2 + v^2 k^2 x (1-x)}} \coth \left( \frac{\sqrt{ m^2 + v^2 k^2 x (1-x)}}{2 T} \right)
\end{equation}
For the fermionic contributions, we include the contribution of the $g_\pm$ fermions with effective mass $m^\ast$ and Fermi
wavevector $k_F$. Calculation of the fermion compressibility yields
\begin{eqnarray}
D_{1f} (k, \epsilon_n) &=& 2 \int \frac{d^2 q}{4 \pi^2}  \frac{(n_F(\varepsilon_{{\bf q} - {\bf k}/2}) -
n_F(\varepsilon_{{\bf q} + {\bf k}/2}))
 }{(i \epsilon_n + {\bf k} \cdot {\bf q}/m^\ast)} \nonumber \\
&\approx &
\frac{m^\ast}{\pi},
\end{eqnarray}
where $n_F$ is the Fermi function.
For the transverse propagator, we obtain from the computation of the fermion current correlations
\begin{eqnarray}
&& D_{2f} (k, \epsilon_n) + \frac{\epsilon_n^2}{k^2} D_{1f} (k, \epsilon_n) \nonumber \\
&&~~~~~= \frac{k_F^2}{2 \pi m^\ast} -
\frac{2}{m^{\ast 2}}  \int \frac{d^2 q}{4 \pi^2} \left( q^2 - \frac{({\bf q} \cdot {\bf k})^2}{k^2} \right)
\frac{(n_F(\varepsilon_{{\bf q} - {\bf k}/2}) -
n_F(\varepsilon_{{\bf q} + {\bf k}/2}))
 }{(i \epsilon_n + {\bf k} \cdot {\bf q}/m^\ast)}
\nonumber \\
&&~~~~~
\approx \frac{k_F |\epsilon_n |}{\pi k }
\end{eqnarray}
Putting all this together, we have the final form of the propagators. The propagator of $A_\tau$ is
\begin{equation}
\frac{1}{N D_{1b} (k) + m^\ast/\pi}
\end{equation}
while that of $A_i$ is
\begin{equation}
\left( \delta_{ij} - \frac{k_i k_j}{k^2} \right) \frac{1}{N D_{2b} (k) + k_F |\epsilon_n|/(\pi k)
}.
\end{equation}


\subsection{Eliashberg equations}

We now address the pairing instability of the $g_{\pm}$ fermions. Both the longitudinal and transverse photons contribute
an attractive interaction between the oppositely charged fermions, which prefers a simple $s$-wave pairing.
However, we also know that the transverse photons destroy the fermionic quasiparticles near the Fermi surface, and so
have a depairing effect. The competition between these effects can be addressed in the usual Eliashberg framework. \cite{scalapino}
Based upon arguments made in Refs.~\onlinecite{rainer,msv}, we can anticipate that the depairing and pairing effects
of the transverse photons exactly cancel each other in the low-frequency limits, because of the $s$-wave pairing. The higher
frequency photons yield a net pairing contribution below a critical temperature $T_c$ which we compute below.

Closely related computations have been carried out by Chubukov and Schmalian \cite{cstc} on a generalized model
of pairing due to the exchange of a gapless bosonic collective mode; our numerical results for $T_c$ below
agree well with theirs, where the two computations can be mapped onto each other. \cite{andrey}

\begin{figure}
\includegraphics[width=4.0 in]{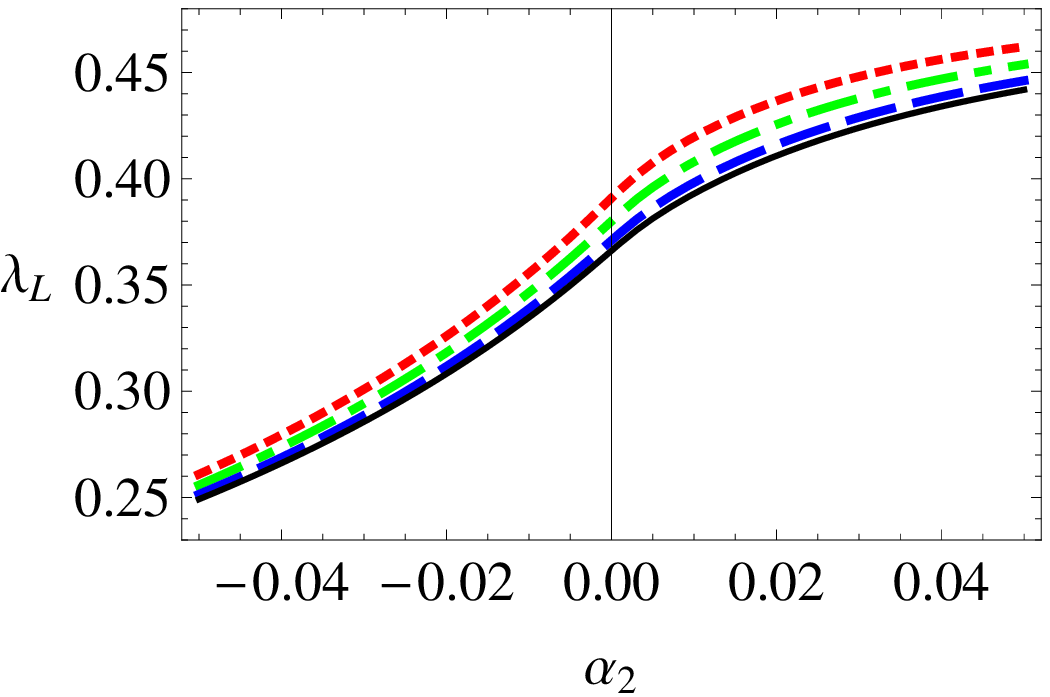}
\includegraphics[width=4.0 in]{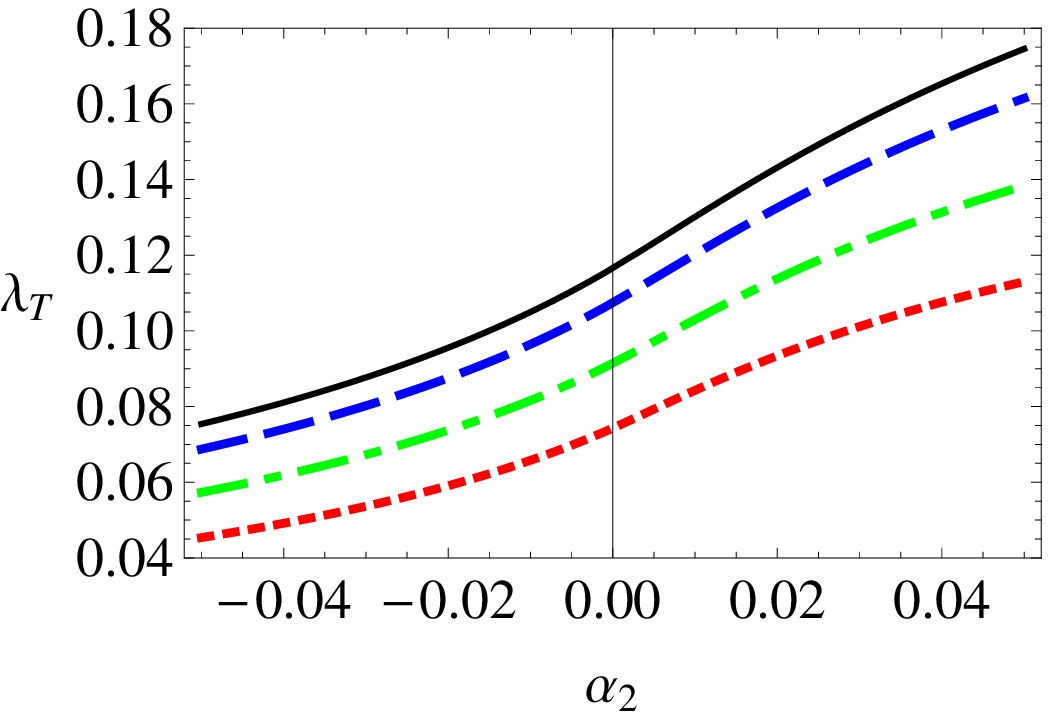}
\caption{(Color online) The pairing coupling constants associated with the longitudinal ($\lambda_L$) and transverse ($\lambda_T (\omega_n) $) gauge
interactions. The parameter $\alpha_2$ measures the distance from the SDW ordering transition in the metal, as defined in
Eq.~(\ref{alpha2}). The dotted (red), dot-dashed (green), dashed (blue), and continuous (black) lines 
correspond to ${\alpha_{1}^2}/{2} = {E_F}/{( m^\ast v^2 )} = $ $0.16, 0.21, 0.26, 0.29$. 
We show  $\lambda_T(\omega_{n} = 8 \pi T)$ with $T/(m^\ast v^2) = 0.016$ for the transverse interaction. 
Note that $\lambda_T (\omega_n) $ function is analytic near $\alpha_2 \sim 0$ in the magnified scale.}
\label{coupling}
\end{figure}

The Eliashberg approximation starts from writing the fermion Green function using Nambu spinor notation.
\begin{eqnarray}
\hat \Sigma (  \omega_{n}) & = & i \omega_{n}(1- Z( \omega_{n})) \hat \tau_{0} + \epsilon \hat \tau_{3} + \phi.( \omega_{n}) \hat \tau_{1} \\
G^{-1}(\epsilon,  \omega_{n})& = & i \omega_{n} Z( \omega_{n}) \hat \tau_{0} - \epsilon \hat \tau_{3} - \phi( \omega_{n}) \hat \tau_{1} \nonumber
\end{eqnarray}
where $\hat{\tau}$ are the Pauli matrices in the particle-hole space.
Then self-consistency equation is constructed by evaluating the self-energy with the above Green function, which 
yields the following equation:
\begin{eqnarray}
 \hat \Sigma ( i \omega_{n})  &=&  T \sum_{\omega_m} \int \frac{d^2 k'}{4 \pi^2} \hat G(k',  \omega_{m}) \widetilde {\rm D}(\vec q, \vec k,  \omega_{m}-  \omega_{n}) \\
& =&  T \sum_{\omega_m} \lambda^{\rm tot} (\omega_{m}-\omega_{n}) \int d \epsilon' \hat G(\epsilon',  \omega_{m}) \nonumber
\end{eqnarray}
Note that the first line is a formal expression, with $ \widetilde {\rm D}(\vec q, \vec k, i \omega_{m})$ being a 
combination of the photon propagator and the matrix elements of the vertex.
The equations are therefore characterized by the coupling $\lambda^{\rm tot} (\omega_n)$; computation of the photon
contribution yields the explicit expression \cite{bmn,ady}
\begin{eqnarray}
\lambda^{\rm tot} (\omega_n) &=& \lambda_T (\omega_n) + \lambda_L \\
\lambda_T (\omega_n) &=& \frac{k_F}{2 \pi^2 m^\ast } \int_0^{2 k_F} dk \frac{\sqrt{1-(k/2k_F)^2}}{
N D_{2b} (k) + k_F |\omega_n|/(\pi k)}  \nonumber \\
\lambda_L &=& \frac{m^\ast}{2 \pi^2 k_F} \int_0^{2 k_F} \frac{dk}{\sqrt{1-(k/2k_F)^2}} \left[
 \frac{1}{N D_{1b} (k) + m^\ast/\pi} \right]
\end{eqnarray}
We have divided the total coupling into two pieces based on the different frequency 
dependence of the longitudinal and transversal gauge boson propagators. The frequency independent term will need a cutoff for the 
actual calculation as we will see below.  The typical behaviors of the dimensionless couplings $\lambda_T(\omega_{n}), \lambda_L$ are
shown in Fig~\ref{coupling}. 

The longitudinal coupling $\lambda_L$ is around $0.35$, and has a significant dependence upon $\alpha_2$, which is a measure
of the distance from the SDW ordering transition. Note that $\lambda_L$ is {\em larger\/} in the SDW-disordered phase ($\alpha_2 > 0$): 
this is a consequence
of the enhancement of gauge fluctuations in this regime. This will be the key to the competing order effect we are looking for:
gauge fluctuations, and hence superconductivity, is enhanced when the SDW order is suppressed.

The transverse gauge fluctuations yield the frequency dependent coupling $\lambda_T (\omega_n)$. This is divergent at
low frequencies with\cite{bmn,ady} $\lambda_T (\omega_n) \sim |\omega_n|^{-1/3}$. As we noted earlier, this divergent
piece cancels out between the normal and anomalous contributions to the fermion self energy. \cite{rainer,msv} 
We plot the dependence
of $\lambda_T (\omega_n)$ on the coupling $\alpha_2$ for a fixed $\omega_n$ in Fig.~\ref{coupling}. As was
for the case of the longitudinal coupling, the transverse contribution is larger in the SDW-disordered phase.

The full self-consistent Eliashberg equations are obtained by matching the coefficients of the Pauli matrices term by term.
\begin{eqnarray}
i \omega_{n} (1- Z( \omega_{n})) &=& - \pi T \sum_{\omega_m } \lambda^{\rm tot} (\omega_m- \omega_n) \frac{ i \omega_{m}  }{\sqrt{\omega_{m}^2  + \Delta^2( \omega_m)}}\\
\Delta( \omega_{n}) &=& \pi T \sum_{\omega_m } \lambda^{\rm tot} (\omega_m-\omega_n) \frac{  \Delta(i \omega_{m})}{\sqrt{\omega_{m}^2  + \Delta^2( \omega_m)}}
\end{eqnarray}
where $\Delta (\omega_n)$ is the frequency-dependent pairing amplitude.

Now we can solve the self-consistent equations to determine the boundary of the superconducting
phase. Our goal is to look for the critical temperature and magnetic field, and we can linearize the equations in $\Delta (\omega_n)$ in these cases; in other words we would neglect the gap functions in the denominator.
\begin{eqnarray}
Z(\omega_n) & = & 1 + \frac{\pi T}{\omega_n} \sum_{\epsilon_n} \mbox{sgn} (\epsilon_n)  \lambda_T (\omega_n - \epsilon_n)   \nonumber \\
&=& 1 + \frac{\pi T}{|\omega_n|} \sum_{|\epsilon_n| < |\omega_n|}  \lambda_T (\epsilon_n)
\end{eqnarray}
\begin{figure}
\includegraphics[width=4.0 in]{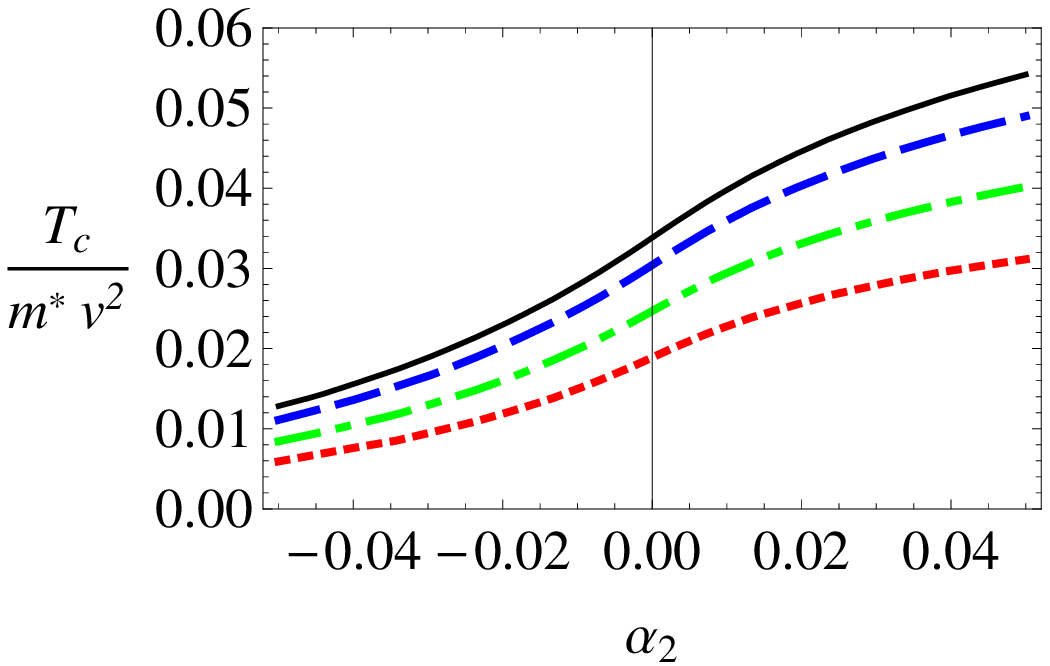}
\includegraphics[width=4.0 in]{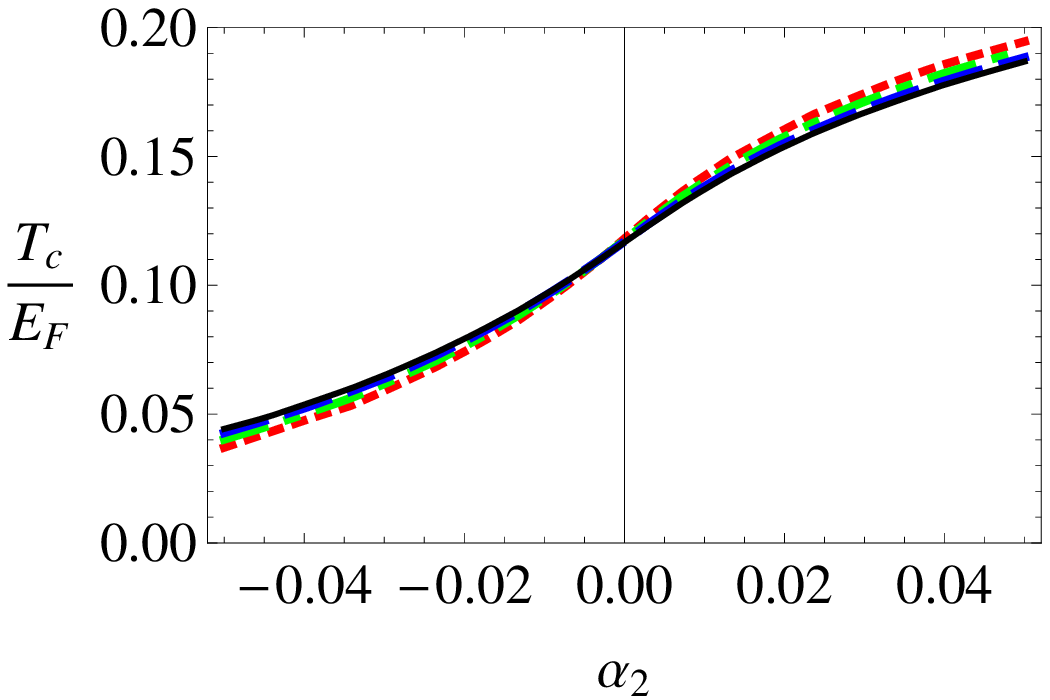}
\caption{(Color online) The critical temperature for superconductivity obtained by solution of the Eliashberg equations. The lines are for the same
parameter values as in Fig.~\ref{coupling}. The top plot has critical temperature scaled with $m^\ast v^2 $, and the bottom is one scaled with $E_F$.}
\label{figTc}
\end{figure}
Then the solution of the critical temperature of linearized Eliashberg equation is equivalent to the condition that the matrix $K (\omega_n, \omega_m)$ first has a positive eigenvalue, where
\begin{equation}
K (\omega_n, \omega_m) =  \lambda_T (\omega_n - \omega_m) + \lambda_L \, \Lambda (\omega_n - \omega_m)  - \delta_{n,m} \frac{ |\omega_n| Z(\omega_n)}{\pi T} \label{kmat}
\end{equation}
with the soft cutoff function with cutoff $E_F$
\begin{equation}
\Lambda (\omega_n) \equiv \frac{1}{1 + c_1 ( \omega_n/E_F)^2}
\end{equation}
where $c_1$ is a constant of order unity. The cutoff $E_F$ is the highest energy scale of the electronic structure, so it is not unnatural to set the cutoff with the scale. With this, the numerics is well-defined and we plot the resulting critical temperature in Fig.~\ref{figTc}.

\begin{figure}
\includegraphics[width=4.0 in]{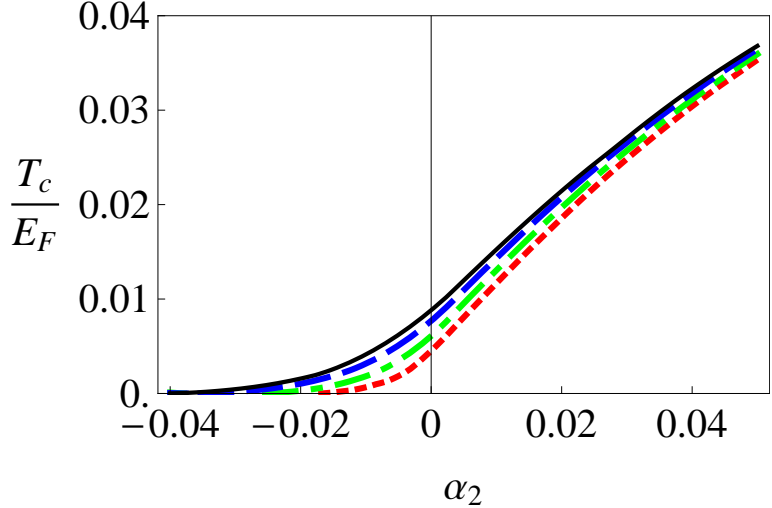}
\caption{(Color online) As in the top panel of Fig.~\ref{figTc}, but with only the transverse pairing interaction, $\lambda_T ( \omega_n )$,
included.}
\label{TcTrans}
\end{figure}

For comparison, we show in Fig.~\ref{TcTrans} 
the results for $T_c$ obtained in a model with only the transverse interaction associated with $\lambda_T (\omega_n)$.
We can use this $T_c$ to define an effective transverse coupling, $\overline{\lambda}_T$, by $T_c /E_F = \exp (-1 /\overline{\lambda}_T)$.
Using $T_c / E_F \approx 0.008$ for $\alpha_2 \approx 0$ in Fig.~\ref{TcTrans}, we obtain $\overline{\lambda}_T \approx 0.2$.
This is of the same order as the longitudinal coupling $\lambda_L$ for  $\alpha_2 \approx 0$ in Fig.~\ref{coupling}.

Bigger attractive interactions $\lambda_T (\omega_n) $ and $ \lambda_L$ clearly induces a higher critical temperature in the SDW-disordered region.  Note that this behavior is  different from the one of previous SDW-mediated superconductivity.\cite{dwave} (See the results of Ref .~\onlinecite{abanov} Fig. [4]; near the critical region, $T_c$ shows the opposite behavior there.) 
We have also compared the plots obtained by
scaling $T_c$ by $m^\ast v^2$ and $E_F$. The dependencies on the parameter $\alpha_1$ are reversed in two plots in the 
SDW-disordered region. To interpret $\alpha_1$ as the doping related parameter, we should choose the scaling by $m^\ast v^2$ because the mass $m^\ast$ and spin wave velocity $v$ are not affected much by doping. With this scaling (the first plot in Fig.~\ref{figTc}), the critical temperature rises with increase doping at fixed $\alpha_2$; of course, in reality, $\alpha_2$ is 
also an increasing function of doping.
	
\subsection{Critical field}

This subsection will extend the above analysis to compute the upper-critical magnetic field, $H_{c2}$ at $T=0$. 
We will neglect the weak Zeeman coupling of the applied field, and assume that it couples only to the 
orbital motion of the $g_\pm$ fermions. This means that $\mathcal{L}_g$ in Eq.~(\ref{lg}) is modified to
\begin{eqnarray}
\mathcal{L}_g &=& g_+^\dagger \left[ (\partial_\tau - i A_\tau )  - \frac{1}{2m^*} ( {\bm \nabla} - i {\bf A} -i (e/c) {\bf a}  )^2 - \mu  \right] g_+
\nonumber \\ &+& g_-^\dagger \left[ (\partial_\tau + i A_\tau )  - \frac{1}{2m^*} ({\bm  \nabla} + i {\bf A} - i (e/c) {\bf a} )^2 - \mu \right] g_-
, \label{lgh}
\end{eqnarray}
where ${\bm \nabla} \times {\bm a} = H$ is the applied magnetic field.

Generally, the magnetic field 
induces non-local properties in the Green's function. However, in the vanishing gap limit, 
Helfand and Werthamer proved the non-locality only appears as a phase factor (see Ref. \onlinecite{hw}). 
The formalism has been developed by Shossmann and Schachinger \cite{scha}, and we will follow their method. 
As they showed, in the resulting equation for $H_{c2}$, the magnetic field only appears in the
modification of the frequency renormalization $Z (\omega_n )$. 

The Eliashberg equations in zero magnetic contain a term $|\omega_n Z(\omega_n)|$, which comes
from the inverse of the Cooperon propagator type at momentum $q=0$, $C(\omega_n, 0)$, where
\begin{eqnarray}
C(\omega_n, q) &=& \int \frac{d^2 p}{4 \pi^2} \frac{1}{(-i \omega_n Z(\omega_n) + \varepsilon_{{\bf p} + {\bf q}})
(i \omega_n Z(\omega_n) + \varepsilon_{-{\bf p}})} \nonumber \\
&\approx& N(0) \int_0^{2 \pi} \frac{d \theta}{2 \pi} \int_{-\infty}^{\infty} d \varepsilon
 \frac{1}{(-i \omega_n Z(\omega_n) + \varepsilon + v_F q \cos \theta)
(i \omega_n Z(\omega_n) + \varepsilon)} \nonumber \\
&=& \frac{2 \pi N(0)}{\sqrt{ 4 \omega_n^2 Z^2 (\omega_n) + v_F^2 q^2}}
\end{eqnarray}
where $N(0)$ is the density of states at the Fermi level per spin.

Now we discuss the extension of this to $H=0$, as described in Refs.~\onlinecite{hw,gal,scha}.
For this, we need to replace $C(\omega_n,0)$ by the smallest eigenvalue of the operator
\begin{equation}
\hat{\mathcal{L}} (\omega_n) = \int d^2 \rho \int \frac{d^2 q}{4 \pi^2} C(\omega_n, q) e^{i {\bf q} \cdot {\bm \rho}}
e^{- i {\bm \rho} \cdot \hat{\bm \pi}}
\end{equation}
where $\hat{\bm \pi} = \hat{\bf p} - (2 e/\hbar c) {\bf A} (\hat{\bf r})$.
Using Eq.~(22) from Ref.~\onlinecite{gal}, we find the smallest eigenvalue of $\hat{\mathcal{L}} (\omega_n)$ is
\begin{eqnarray}
\mathcal{L}_0 (\omega_n) &=& \int_0^\infty \rho d \rho \int_0^{\infty} q dq J_0(q\rho)
C(\omega_n, q) e^{- \rho^2 /(2 r_H^2)} \nonumber \\
&=&  r_H^2 \int_0^\infty q dq e^{-q^2 r_H^2 /2} C(\omega_n,q)
\end{eqnarray}
where $r_H = \sqrt{\hbar c/2eH}$ is the magnetic length.

So the only change in the presence of a field is that the wavefunction renormalization $Z(\omega_n)$
is replaced by $Z_H (\omega_n)$, where
\begin{eqnarray}
\frac{1}{Z_H (\omega_n)}
&=&  2 |\omega_n| r_H^2 \int_0^\infty q dq \frac{e^{-q^2 r_H^2 /2}}{\sqrt{4 \omega_n^2 Z^2 (\omega_n) + v_F^2 q^2}} \\
&=&  2 |\omega_n|  \int_0^\infty x dx \frac{e^{-x^2  /2}}{\sqrt{4 \omega_n^2 Z^2 (\omega_n) + v_F^2 r_H^{-2} x^2}}
\label{wavefunction}
\end{eqnarray}

\begin{figure}
\includegraphics[width=4.0 in]{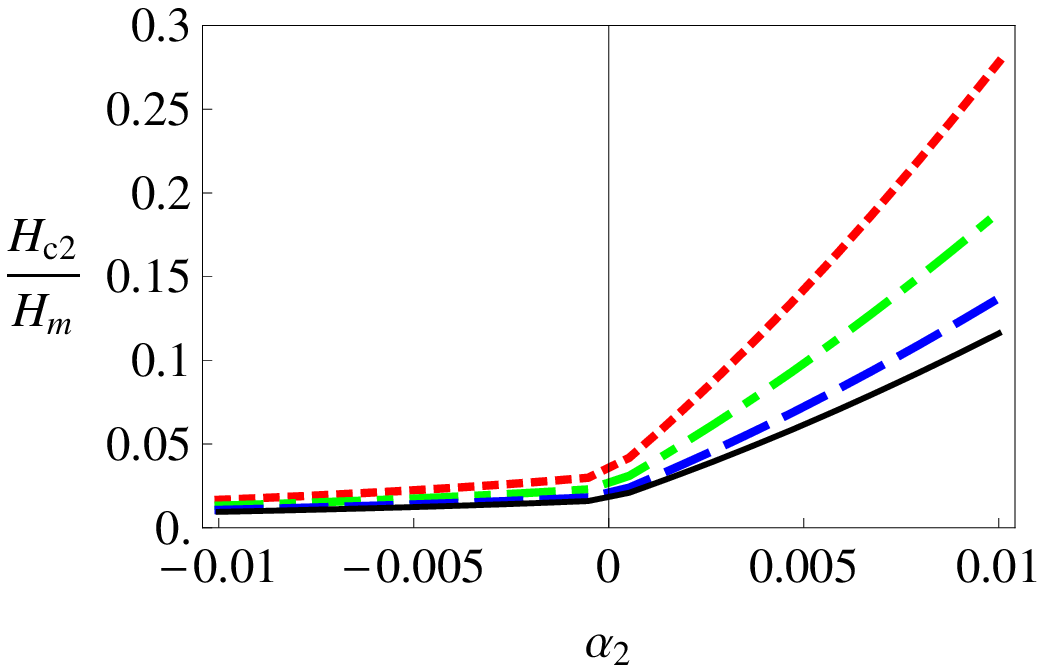}
\includegraphics[width=4.0 in]{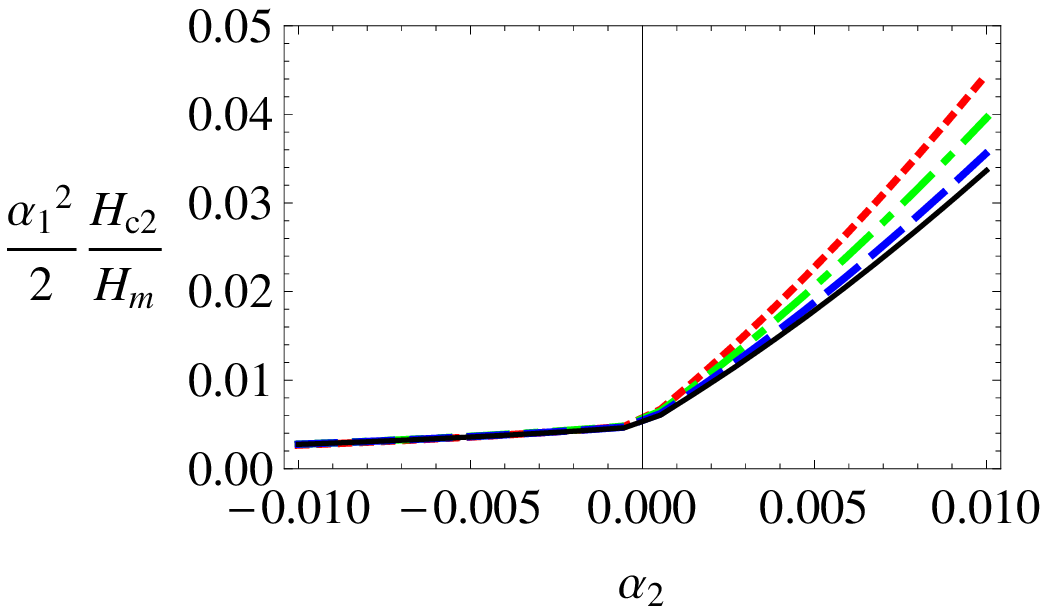}
\caption{(Color online) The upper critical field $H_{c2}$ as a function of $\alpha_1$ and $\alpha_2$ using the same conventions
as in Fig.~\ref{coupling}. The magnetic field is measured with the units induced by the fermion mass via $H_m$ defined in
Eq.~(\ref{hmdef}). }
\label{figHc}
\end{figure}
We can now insert the modified $Z (\omega_n)$ into Eq.~(\ref{wavefunction}) into Eq.~(\ref{kmat}), and so compute $H_{c2}$ as a function of both $\alpha_1$
and the SDW tuning parameter $\alpha_2$. The natural scale for the magnetic field is
\begin{equation}
H_m \equiv \left( \frac{\hbar c}{2 e} \right) k_F^2 \approx 534~{\rm Tesla}, \label{hmdef}
\end{equation}
where in the last step we have used values from the quantum oscillation experiment \cite{doiron} quoted
in Section~\ref{sec:intro}. 
Our results for $H_{c2}/H_m$ are shown in Fig.~\ref{figHc}.
We can see that the critical field dependence on $\alpha_{2}$ is similar to the critical temperature dependence: 
it is clear that SDW competes with superconductivity, and that $H_{c2}$ decreases as the SDW ordering is enhanced
by decreasing $\alpha_2$.
Also, we can compare this with the phenomenological phase diagram of Fig \ref{figdemler}; the critical field line in Fig.~\ref{figHc} 
determines the line B-M-D  within Eliashberg approximation. Finally, the values of $H_{c2}$ in Fig.~\ref{figHc} are quite compatible with the quantum oscillation experiments. \cite{doiron,cooper,nigel,cyril,suchitra,louis}

\section{Shift of SDW ordering by superconductivity}
\label{sec:shift}

We are interested in the feedback on the strength of magnetic order
due to the onset of superconductivity. Rather than using a self-consistent approach,
we will address the question here systematically in a $1/N$ expansion.

We will replace the fermion action in Eq.~(\ref{lg}) by
a theory which  has $N/2$ copies of the electron pockets
\begin{eqnarray}
\mathcal{L}_g &=& \sum_{a=1}^{N/2} \Biggl\{ g_{+a}^\dagger \left[ (\partial_\tau - i A_\tau )  - \frac{1}{2m^*} ( {\bm \nabla} - i {\bf A}  )^2 - \mu  \right] g_{+a}
\nonumber \\ &~&~~~+ g_{-a}^\dagger \left[ (\partial_\tau + i A_\tau )  - \frac{1}{2m^*} ({\bm  \nabla} + i {\bf A} )^2 - \mu \right] g_{-a} \nonumber \\
&~&~~~~~~~~~~~~ - \Delta \, g_{+ a} g_{-a} - \Delta \, g_{-a}^\dagger g_{+a}^\dagger \Biggr\}
\end{eqnarray}
Here we consider the gauge boson fluctuation more rigorously in the sense of accounting for full fermion and boson polarization functions.
But we will treat the fermion pairing amplitude $\Delta$ as externally given: the previous
section described how it could be determined in the Eliashberg theory with approximated polarization.

The large $N$ expansion proceeds by integrating out the $z_\alpha$ and the $g_{\pm a}$, and then
expanding the effective action for $\varrho$ and $A_\mu$ -- formally this has the same structure as
the computation in Ref.~\onlinecite{ribhu}, generalized here to non-relativistic fermions.
At $N=\infty$, the $g_{\pm a}$ and $z_{\alpha}$ remain decoupled because the gauge propagator is 
suppressed by a prefactor of ${1}/{N}$. 
So at this level, the magnetic critical point is not affected by the presence of the fermions, and appears at
$t=t_c^0$ where
\begin{equation}
\frac{1}{t_c^0} = \int \frac{d \omega d^2 k}{8 \pi^3} \frac{1}{\omega^2 + v^2 k^2}.
\end{equation}

\begin{figure}
\includegraphics[width=4.0 in]{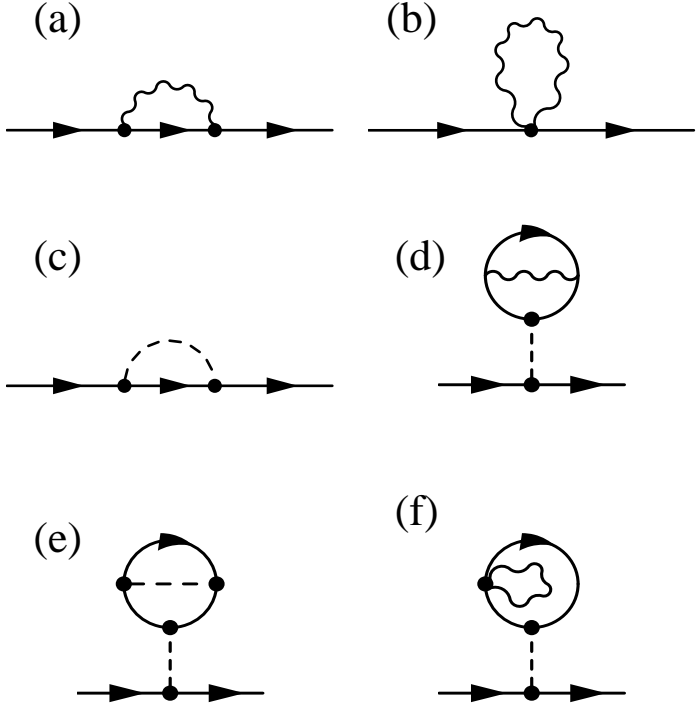}
\caption{Feynman diagrams for the self energy of $z_\alpha$ from Ref.~\onlinecite{ribhu}. 
The full line represents $z_\alpha$, the wavy line
is the $A_\mu$ propagator, and the dashed line is the $\varrho$ propagator which imposes the
length constraint on $z_\alpha$.}
\label{figribhu}
\end{figure}

We are interested in determining the $1/N$ correction to the magnetic quantum critical point, which
we write as
\begin{equation}
\frac{1}{t_c (\Delta) } = \frac{1}{t_c^0} + \frac{1}{N} F( \Delta);
\end{equation}
note that in the notation of Fig.~\ref{figdemler}, $t_c \equiv t_c (\Delta)$.
The effect of superconductivity on the magnetic order will therefore be determined by
$F(\Delta) - F(0)$, which is the quantity to be computed.
The shift of the critical point at this order will be determined by the graphs in
Fig. 3 of Ref.~\onlinecite{ribhu}, which are reproduced here in Fig.~\ref{figribhu}. 
Evaluating these graphs we find
\begin{eqnarray}
F(\Delta) &=& \int \frac{d^2 q d \omega}{8 \pi^3}  \int \frac{d^2 p d \epsilon}{8 \pi^3} \frac{1}{(\epsilon^2 + v^2 p^2)^2} \Biggl[
\frac{1}{D_1 (q, \omega)} \left( \frac{(2 \epsilon + \omega)^2}{(\epsilon+\omega)^2 + v^2 (p + q)^2} - \frac{\omega^2}{\omega^2 + v^2 q^2} \right) \nonumber \\
&~&~~~~~~~~~~~~~~~~~+
\frac{1}{[D_2 (q, \omega) + (\omega^2/q^2) D_1 (q, \omega)]}  \left(\frac{4 v^4 (p^2 - (p.q)^2/q^2)}{(\epsilon+\omega)^2 + v^2 (p + q)^2} \right)  \nonumber \\
&~&~~~~~~~~~~~~~~~~~
+ \frac{1}{\Pi_{\varrho} (q, \omega)}  \left( \frac{1}{\omega^2 + v^2 q^2} - \frac{1}{(\omega+ \epsilon)^2 + v^2 (p+q)^2} \right)
\Biggr], \label{e2}
\end{eqnarray}
where $1/\Pi_\varrho (q, \omega) = 8 \sqrt{\omega^2 + v^2 q^2}$ is the propagator of the Lagrange multiplier field $\varrho$. The last term involving $\Pi_\varrho$ is independent of $\Delta$, and so will drop out of our final expressions measuring 
the influence of superconductivity: we will therefore omit this term in subsequent expressions for $F (\Delta)$. 

It is now possible to evaluate the integrals over $p$ and $\epsilon$ analytically. This is done by using
a relativistic method in 3 spacetime dimensions. Using a 3-momentum notation in which $P_\mu \equiv (v p_i, \epsilon)$ and
$Q_\mu \equiv (v q_i, \omega)$ and $\int_P \equiv v^2 \int d \epsilon d^2 p/(8 \pi^3)$,
some useful integrals obtained by dimensional regularization are:
\begin{eqnarray}
\int_P \frac{1}{P^4} &=& 0 \nonumber \\
\int_P \frac{1}{P^4 (P+Q)^2} &=& 0 \nonumber \\
\int_P \frac{P_\mu} {P^4 (P+Q)^2} &=& - \frac{Q_\mu}{16Q^3} \nonumber \\
\int_P \frac{P_\mu P_\nu}{P^4 (P+Q)^2} &=& \frac{\delta_{\mu\nu}}{32 Q} + \frac{Q_\mu Q_\nu}{32 Q^3}.
\end{eqnarray}
While some of the integrals above appear infrared divergent, there are no infrared divergencies in the complete original expression
in Eq.~(\ref{e2}), and we have verified that dimensional regularization does indeed lead to the correct answer obtained from a
more explicit subtraction of the infrared singularities.
Using these integrals, we obtain from Eq.~(\ref{e2})
\begin{eqnarray}
F(\Delta) &=& \int \frac{d^2 q d \omega}{8 \pi^3} \frac{q^2}{8 (\omega^2 + v^2 q^2)^{1/2}} \Biggl[
\frac{1}{(\omega^2 + v^2 q^2) D_1 (q, \omega) }
+
\frac{1}{q^2 D_2 (q, \omega) + \omega^2 D_1 (q, \omega)} \Biggr]. ~~~\label{e3}
\end{eqnarray}
The above expression was obtained in the Coulomb gauge, but we have verified that it is indeed gauge invariant.

We can now characterize the shift of the critical point in the superconductor
by determining the spinon gap, $\Delta_z$, at the coupling $t=t_c (0)$ where there is
onset of magnetic order in the metal {\em i.e.} the spinon gap in the superconductor
at $H=0$ at the value of $t$ corresponding to the line CM in Fig.~\ref{figdemler}.
To leading order in $1/N$, this is given by
\begin{eqnarray}
\frac{\Delta_z}{m^\ast v^2} &=& \frac{4 \pi}{m^\ast} \left( \frac{1}{t_c (\Delta)} - \frac{1}{t_c (0)} \right) \nonumber \\
&=& \frac{4 \pi}{m^\ast N} \left( F (\Delta) - F(0) \right). \label{deltaz}
\end{eqnarray}
This expression encapsulates our main result on the backaction of the superconductivity of the $g_\pm$ fermions, with pairing gap $\Delta$,
on the position of the SDW ordering transition.

Before we can evaluate Eq.~(\ref{deltaz}), we need the gauge field propagators $D_{1,2}$. 
For completeness, we give explicit expressions for the boson and fermionic contributions by writing
\begin{eqnarray}
D_{2}(q,\omega)+\frac{\omega^2}{q^2}D_{1}(q,\omega) &=& D^b_{T}(q,\omega)+D^f_{T}(q,\omega) \\
D_{1}(q,\omega) &=& D^b_{L}(q,\omega)+D^f_{L}(q,\omega).
\end{eqnarray}
We can read off the bosonic polarization functions $D^b_{L,T}(q,\omega)$ from the exact relativistic result of Ref. \onlinecite{ribhu}, and the Eq. (\ref{sa}).
\begin{eqnarray}
D^b_{T}(q,\omega) &=& \frac{\sqrt{v^2 q^2 +\omega^2}}{16} \\
D^b_{L}(q,\omega) &=& \frac{1}{16} \frac{q^2}{\sqrt{v^2 q^2+\omega^2}}
\end{eqnarray}
For the fermion contribution, let us introduce the Nambu spinor Green's function
\begin{eqnarray}
\bar g(q,\omega) & = & \frac{1}{(i \omega)^2-E_q^2} \left(  \begin{array}{cc}
 i \omega+ \xi_q & - \Delta \\
 - \Delta & i \omega- \xi_q  \end{array} \right)  \\
 & =& \int \frac{d \Omega}{\pi} \frac{1}{\Omega- i \omega } {\rm Im}  \left[ \bar g (q,\Omega) \right] \\
{\rm Im} \left[ \bar g (q,\Omega) \right] &=& \frac{(-\pi)}{2 E_q}(\delta(\Omega-E_q)-\delta(\Omega+E_q)) \left(  \begin{array}{cc}
 \Omega+ \xi_q & - \Delta \\
 - \Delta & \Omega- \xi_q  \end{array} \right),
\end{eqnarray}
where $\xi_q = q^2/(2 m^\ast) - \mu$ and $E_q = \sqrt{\xi_q^2 +\Delta^2} $. With the matrix elements of longitudinal and transverse parts, the polarizations of the fermions are as:
\begin{eqnarray}
D^f_{L}(q,\omega) &=& -\int \frac{d^2 k}{(2 \pi)^2} \frac{d \epsilon}{(2 \pi)} {\rm tr} \left[ \bar g(k, \epsilon ) \bar g (q+k,\omega +\epsilon)  \right] \nonumber \\
 &=& \int \frac{d^2 k}{(2 \pi)^2} \int \frac{d \Omega'}{ \pi} \frac{d \Omega}{\pi} \frac{n_F(\Omega')-n_F(\Omega)}{i \omega +\Omega -\Omega'} {\rm tr} \left[ {\rm Im}\bar g(k, \Omega ) {\rm Im}\bar g (q+k,\Omega' )  \right] \nonumber \\
  &=& \int \frac{d^2 k}{(2 \pi)^2}  \frac{1}{2} \left(1-\frac{\xi_k \xi_{k+q}+\Delta^2}{E_k E_{k+q}} \right)\left( \frac{2(E_k+E_{k+q})}{\omega^2+(E_k+E_{k+q})^2} \right).
\end{eqnarray}
\begin{eqnarray}
D^f_{T}(q,\omega) &=& D^f_{T,dia} + D^f_{T,para} \nonumber \\
D^f_{T,para} &=& - \int \frac{d^2 k}{(2 \pi)^2}\frac{1}{(m^\ast)^2} \left(k^2 - \frac{(k \cdot q)^2}{q^2} \right) \int {\frac{d \epsilon}{(2 \pi)} \rm tr} \left[ \bar g(k, \epsilon ) \hat \tau_3 \bar g (q+k,\omega +\epsilon) \hat \tau_3 \right] \nonumber \\
 &=& -\int \frac{d^2 k}{(2 \pi)^2}\frac{k^2 {\rm sin}^2 \theta}{(m^\ast)^2}  \int \frac{d \Omega}{ \pi} \frac{d \Omega'}{\pi}
 \frac{n_F(\Omega')-n_F(\Omega)}{i \omega +\Omega -\Omega'}  \nonumber  \nonumber \\
 &~&~~~~~~~~~~~~~~~~~~~~~ \times {\rm tr} \left[ {\rm Im }\bar g(k, \Omega ) \hat \tau_3 {\rm Im } \bar g (q+k,\Omega') \hat \tau_3 \right] \nonumber \\
  &=& -\int \frac{d^2 k}{(2 \pi)^2}\frac{k^2 {\rm sin}^2 \theta}{(m^\ast)^2} \frac{1}{2} \left(1-\frac{\xi_k \xi_{k+q} -\Delta^2}{E_k E_{k+q}}\right)\left( \frac{2(E_k+E_{k+q})}{\omega^2+(E_k+E_{k+q})^2} \right)\\
D^f_{T,dia} &=& \frac{\rho_f}{m^\ast} ,
 \end{eqnarray}
where $\rho_f$ is the density of the fermions and $\hat{\tau}_3$ is a Pauli matrix in the Nambu particle-hole space. With these results we are now ready to evaluate Eq.~(\ref{deltaz}).

\begin{figure}
\includegraphics[width=4.0 in]{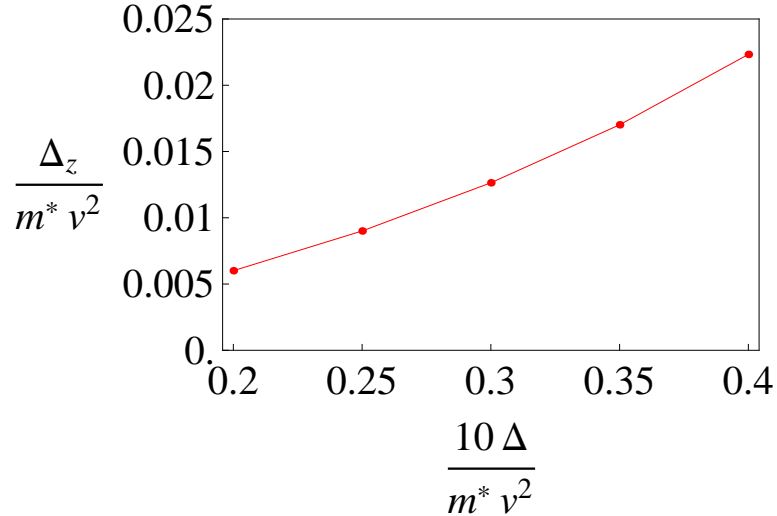}
\caption{The energy $\Delta_z$ in Eq.~(\ref{deltaz}) determining the value of the shift in the
SDW ordering critical point, $t_c (0) - t_c ( \Delta ) $. 
The horizontal axis is the externally given superconducting gap. For numerics we fix the parameter $ \alpha_1 /2= {E_F}/{m^\ast v^2} =0.3$ }
\label{spinongap}
\end{figure}

One of the key features of the theory of competing orders was the enhanced stability of SDW ordering
in the metallic phase. This corresponds to feature B discussed in Section~\ref{sec:intro}: $t_c (0) > t_c $ in Fig.~\ref{figdemler}.
In the notation of our key result in Eq.~(\ref{deltaz}), where $t_c (\Delta) \equiv t_c$, this requires $\Delta_z > 0$. 
We show numerical evaluations of Eq.~(\ref{deltaz}) in Fig~\ref{spinongap} and find this indeed the case.
(The values of $\Delta$ used in Fig.~\ref{spinongap} are similar to those obtained in Section~\ref{sec:eliash}
near the SDW ordering critical point.)
Indeed, the sign of $\Delta_z$ is easily understood. In the metallic phase, the gauge fluctuations are quenched by
excitations of the Fermi surface. On the other hand, in the superconducting state, this effect is no longer
present: gauge fluctuations are enhanced and hence SDW ordering is suppressed. Note that the fact that the $g_\pm$ fermions
have opposite gauge charges is crucial to this conclusion. The ordinary Coulomb interaction, under which the $g_\pm$ have
the same charge, continues to be screened in the superconductor. In contrast, a gauge force which couples with opposite
charges has its polarizability strongly suppressed in the superconductor, much like the response of a BCS superconductor to a Zeeman field.

\section{Conclusions}
\label{sec:conc}

This paper has described the phase diagram of a simple 'minimal model' of the underdoped, hole-doped
cuprates contained in Eqs.~(\ref{m1}), (\ref{lg}), and (\ref{lz}). This theory describes bosonic neutral spinons $z_\alpha$ and
spinless charge $-e$ fermion $g_\pm$ coupled via a U(1) gauge field $A_\mu$. We have shown that the theory reproduces key
aspects of a phenomenological phase diagram \cite{demler,ssrmp} of the competition between SDW order and superconductivity in Fig.~\ref{figdemler}
in an applied magnetic field, $H$. This phase diagram has successfully predicted a number of recent experiments, as was discussed
in Section~\ref{sec:intro}. 

In particular, in Section~\ref{sec:eliash}, we showed that the minimal model had a $H_{c2}$ which 
decreased as the SDW ordering was enhanced by decreasing the coupling $t$ in Eq.~(\ref{lz}). 

Next, in Section~\ref{sec:shift}, we showed
that the onset of SDW ordering in the normal state with $H>H_{c2}$ occurred at a value $t=t_c (0)$ which was distinct
from the value $t=t_c (\Delta)$ in the superconducting state with $H=0$. As expected from the competing order picture in Fig.~\ref{figdemler},
we found $t_c (0) > t_c (\Delta)$. The enhanced stability of SDW ordering in the metal was a consequence of the suppression of $A_\mu$
gauge fluctuations by the $g_\pm$ Fermi surfaces. These Fermi surfaces are absent in the superconductor, and as a result the gauge
fluctuations are stronger in the superconductor.

\begin{figure}
\includegraphics[width=4.0in]{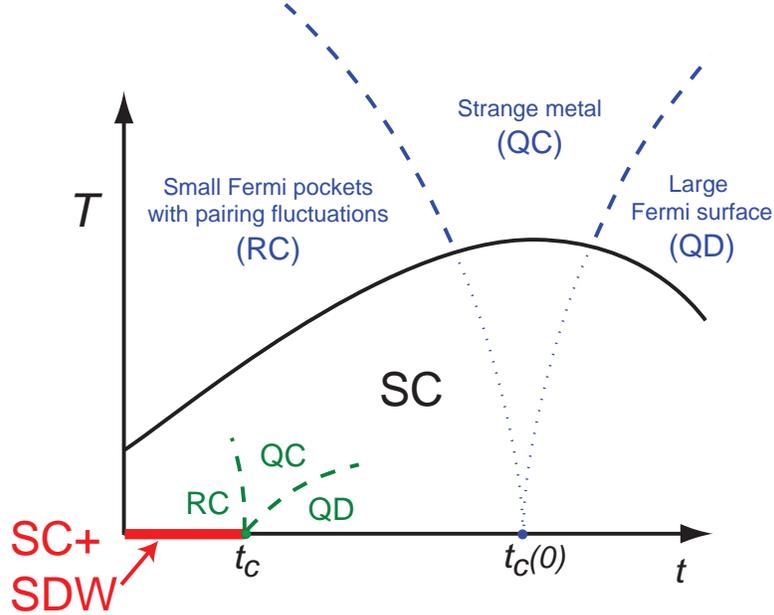}
\caption{Proposed finite temperature crossover phase diagram for the cuprates. 
The labels at $T=0$ are as in Fig.~\ref{figdemler}: the onset of SDW order in the superconductor
is at $t_c \equiv t_c (\Delta)$, while $t_c (0)$ is a `hidden' critical point which can be observed only
at $H > H_{c2}$ as in Fig.~\ref{figdemler}. The computations in Section~\ref{sec:shift} show that
$t_c (0) > t_c (\Delta)$. The full line is the phase transition at $T_c$ representing loss
of superconductivity. The dashed lines are crossovers in the fluctuations of the SDW order.
The dotted lines are guides to the eye and do not represent
any crossovers. Thus, in the pseudogap regime at $T>T_c$ the SDW fluctuations are in the `renormalized classical' \cite{chn}
(RC) regime; upon lowering temperature, they crossover to the `quantum critical' (QC) and  `quantum disordered' (QD) regime in the superconductor.
}
\label{figT}
\end{figure}
We conclude this paper by discussing implications of our results
for the phase diagram at $T>0$, and in particular for the pseudogap regime above $T_c$. 
In our application of the main result in Section~\ref{sec:shift}, $t_c (0) > t_c (\Delta)$ we have assumed that
the $\Delta=0$ state was reached by application of a magnetic field. However, this result also applies if $\Delta$
is suppressed by thermal fluctuations above $T_c$. Unlike $H$, thermal fluctuations will also directly affect
the SDW order, in addition to the indirect effect through suppression of superconductivity.  
In particular in two spatial dimensions there can be no long-range SDW order at any $T>0$.  These considerations
lead us to propose the crossover phase diagram in Fig.~\ref{figT} in the $T$, $t$ plane.
We anticipate that $t_c (0)$ is near optimal doping. Thus in the underdoped regime above $T_c$, there is local SDW order
which is disrupted by classical thermal fluctuations: this is the so-called `renormalized classical' \cite{chn} regime of the
hidden metallic quantum critical point at $t_c (0)$. Going below $T_c$ in the underdoped regime, we eventually reach 
the regime controlled by the quantum critical point associated with SDW ordering in the superconductor, which is at $t_c (\Delta)$.
Because $t_c (\Delta) < t_c (0)$, the SDW order can now be `quantum disordered' (QD). Thus neutron scattering in the superconductor
will not display long-range SDW order as $T \rightarrow 0$, even though there is a RC regime of SDW order above $T_c$.
This QD region will have enhanced charge order correlations\cite{vojtass,ssrmp,rkk3,stripes}; this charge order can survive 
as true long-range order below $T_c$, even 
though the SDW order does not. Thus we see that in our theory the underlying competition is between
superconductivity and SDW order, while there can be substantial charge order in the superconducting phase.

Further study of the nature of the quantum critical point at $t_c (0)$ in the metal is an important direction 
for further research. In our present formulation in Eq.~(\ref{m1}), this point is a transition from a conventional metallic
SDW state to an `algebraic charge liquid' \cite{rkk2} in the O(4) universality class.\cite{rkk3}
However, an interesting alternative possibility is a transition directly to the large Fermi surface state.\cite{senthil}

Finally, we note that a number of experimental studies \cite{yazdani,kohsaka1,kohsaka2,louisnernst,eric1,eric2,gross} have discussed a scenario
for crossover in the cuprates which is generally consistent with our Fig.~\ref{figT}.

\acknowledgements

We thank A.~Chubukov, V.~Galitski, P.~D.~Johnson, R.~Kaul, A.~Keren, Yang Qi, L.~Taillefer, Cenke Xu, and A.~Yazdani for valuable discussions.
We are especially grateful to A.~Chubukov for pointing out numerical errors in an earlier version of this paper.
This research was supported by the NSF under grant DMR-0757145, by the FQXi
foundation, and by a MURI grant from AFOSR. E.G.M. is also supported in part by a Samsung scholarship.

\appendix

\section{Field relations from spin density wave theory}
\label{sdw}

This appendix will give a derivation of the relation (\ref{Pgz}) between the physical electron operators
$c_{\alpha}$ and the fields $g_\pm$ and $z_\alpha$ using spin density wave theory. This will complement
the derivation obtained from the doped Mott insulator approach in previous works.  \cite{rkk3,gs}

We begin the quasiparticle Hamitonian which determines the `large' Fermi surface in the overdoped regime
\begin{equation}
H_0 = - \sum_{i<j} t_{ij} c^{\dagger}_{i \alpha} c_{j \alpha} \equiv \sum_{{\bf k}} \varepsilon_{{\bf k}} c^{\dagger}_{{\bf k}\alpha} c_{{\bf k}\alpha}
\end{equation}
where we choose the dispersion $\varepsilon_{{\bf k}}$ to agree with the measured Fermi surface.
In the presence of spin density wave order, $\vec{\varphi}$ at wavevector ${\bf K} = (\pi, \pi)$, 
we have an additional term which mixes electron states with momentum separated by ${\bf K}$
\begin{equation}
H_{\rm sdw} = -\vec{\varphi} \cdot \sum_{{\bf k}, \alpha, \beta} c_{{\bf k}, \alpha}^\dagger \vec{\sigma}_{\alpha\beta}
c_{{\bf k} + {\bf K}, \beta}
\end{equation}
where $\vec{\sigma}$ are the Pauli matrices. 

Now we focus on the electrons which are near the electron pockets.
Let us write 
\begin{equation}
c_{(0,\pi)\alpha} \equiv c_{1\alpha}~~~,~~~ c_{(\pi,0)\alpha} \equiv c_{2 \alpha}
\end{equation}
and $\varepsilon_{(0,\pi)} = \varepsilon_{(\pi,0)} = \varepsilon_0$.
Then, for N\'eel order polarized as $\vec{\varphi} = (0,0,\varphi)$ with $\varphi>0$, the Hamiltonian for these electrons is
\begin{eqnarray}
H_0 + H_{\rm sdw} &=& \varepsilon_0 \left( c^\dagger_{1\alpha} c_{1 \alpha} + c_{2 \alpha}^\dagger
c_{2 \alpha} \right) \nonumber \\ &~&~~~- \varphi  \left( c^\dagger_{1 \uparrow} c_{2 \uparrow} - c_{1 \downarrow}^\dagger c_{2 \downarrow} + c^\dagger_{2 \uparrow} c_{1 \uparrow} - c_{2 \downarrow}^\dagger c_{1 \downarrow} \right) 
\end{eqnarray}
We diagonalize this by writing
\begin{equation}
H_0 + H_{\rm sdw} = (\varepsilon_0 - \varphi)\left( g_+^\dagger g_+ + g_-^\dagger g_- \right)
+ (\varepsilon_0 + \varphi)\left( h_+^\dagger h_+ + h_-^\dagger h_- \right)
\end{equation}
where
\begin{eqnarray} 
c_{1 \uparrow} &=& ( g_+ + h_+)/\sqrt{2} \nonumber \\
c_{2 \uparrow} &=& ( g_+ - h_+)/\sqrt{2} \nonumber \\
c_{1 \downarrow} &=& ( g_- + h_-)/\sqrt{2} \nonumber \\
c_{2 \downarrow} &=& ( -g_- + h_-)/\sqrt{2} 
\end{eqnarray}

Now the main approximation we make here is to neglect the higher energy $h_\pm$ fermions.
We obtain the electron operators for a general polarization of the N\'eel order as in Eq.~(\ref{pz}) by
performing an SU(2) rotation defined by the $z_{\alpha}$ (and dropping the unimportant factor of $1/\sqrt{2}$)
\begin{equation}
\left( \begin{array}{c} c_{1\uparrow} \\ c_{1\downarrow} \end{array} \right) = \mathcal{R}_z \left( \begin{array}{c} g_+ \\ g_- \end{array} \right)~~~~~;~~~~~\left( \begin{array}{c} c_{2\uparrow} \\ c_{2\downarrow} \end{array} \right) = \mathcal{R}_z \left( \begin{array}{c} g_+ \\ -g_- \end{array} \right)
\end{equation}
where the SU(2) rotation is
\begin{equation}
\mathcal{R}_z = \left( \begin{array}{cc}
z_\uparrow & -z_\downarrow^\ast  \\
z_\downarrow &  z_\uparrow^\ast   
 \end{array} \right).
\end{equation}
These results lead immediately to Eq.~(\ref{Pgz}). In the superconducting state, where $\langle g_+ g_- \rangle \neq 0$,
they yield
\begin{eqnarray}
\langle c_{1 \uparrow} c_{1 \downarrow} \rangle &=& \left \langle \left( |z_\uparrow |^2 + |z_\downarrow |^2 \right)
g_+ g_- \right \rangle \nonumber \\
\langle c_{2 \uparrow} c_{2 \downarrow} \rangle &=& -\left \langle \left( |z_\uparrow |^2 + |z_\downarrow |^2 \right)
g_+ g_- \right \rangle, 
\end{eqnarray}
which implies a $d$-wave pairing signature for the electrons.

\end{document}